\documentclass[amsmath,amssymb,11pt]{article}
\usepackage{jheppub2}
\pdfoutput=1
\usepackage{graphicx,epsfig}
\usepackage{float}
\usepackage{subfloat}
\usepackage[utf8]{inputenc}
\usepackage{hyperref}
\usepackage{caption}
\usepackage{mathrsfs}
\usepackage{color}
\usepackage{overpic}
\usepackage[dvipsnames]{xcolor}
\usepackage{physics}
\usepackage{amsmath}
\usepackage{amsthm}
\usepackage[multiple]{footmisc}
\usepackage[]{cleveref}
\usepackage{subfigure}

\theoremstyle{conjecture}

\newcommand{\rH}{r_\text{H}}

\newcommand{\ta}{\tilde a}

\newcommand{\be}{\begin{equation}}
\newcommand{\ee}{\end{equation}}

\usepackage{tikz}

\usetikzlibrary{decorations.pathmorphing}
\usetikzlibrary{decorations.pathreplacing,decorations.markings}

\usetikzlibrary{backgrounds,automata}

\title{Weak cosmic censorship and the rotating quantum BTZ black hole}

\author[a,b]{Antonia M. Frassino,}
\emailAdd{antonia.frassino@uah.es}
\author[c,d,e]{Jorge V. Rocha}
\emailAdd{jorge.miguel.rocha@iscte-iul.pt}
\author[b,f,g]{and Andrea P. Sanna}
\emailAdd{asanna@dsf.unica.it}

\affiliation[a]{Departamento de F\'{i}sica y Matem\'{a}ticas, University of Alcal\'{a}, Campus universitario 28805, Alcal\'a de Henares (Madrid), Spain}                

\affiliation[b]{Departament de F\'{i}sica Qu\'{a}ntica i Astrof\'{i}sica, Institut de Ci\'{e}ncies del Cosmos, Universitat de Barcelona, Mart\'{i} i Franqu\'{e}s 1, E-08028 Barcelona, Spain} 

\affiliation[c]{Departamento de Matem\'atica, ISCTE--Instituto Universit\'ario de Lisboa, Avenida das For\c{c}as Armadas, 1649-026 Lisboa, Portugal}
\affiliation[d]{Centro de Astrof\'isica e Gravita\c{c}\~ao--CENTRA, Instituto Superior T\'ecnico--IST, Universidade de Lisboa--UL, Av. Rovisco Pais 1, 1049-001 Lisboa, Portugal}
\affiliation[e]{Instituto de Telecomunica\c{c}\~oes--IUL, Avenida das For\c{c}as Armadas, 1649-026 Lisboa, Portugal} 

\affiliation[f]{Dipartimento di Fisica, Universit\`a di Cagliari, Cittadella Universitaria, 09042 Monserrato, Italy}
\affiliation[g]{I.N.F.N, Sezione di Cagliari, Cittadella Universitaria, 09042 Monserrato, Italy}

\abstract{Tests of the weak cosmic censorship conjecture examine the possibility of the breakdown of predictivity of the gravitational theory considered, by checking if curvature singularities typically present in black hole spacetimes are concealed within an event horizon at all times.
A possible method to perform such tests was proposed by Wald and consists in trying to overspin an extremal rotating black hole by throwing at it a test particle with large angular momentum.
In this paper, we analyze the effects of dropping a test particle into an extremal quantum rotating BTZ black hole, whose three-dimensional metric captures the exact backreaction from strongly coupled quantum conformal fields. Our analysis reveals that, despite the inclusion of quantum effects, and akin to the classical scenario, these attempts to destroy the black hole are doomed to be unsuccessful. Particles carrying the maximum angular momentum and still falling into an extremal quantum BTZ black hole can, at most, leave it extremal. Nevertheless, we found numerical evidence that large backreaction of the quantum fields tends to disfavor violations of cosmic censorship.}

\begin{document}
\maketitle

\section{Introduction \label{sec:intro}}

Black holes (BHs) in General Relativity, and in many extensions thereof, harbor singularities. These are regions of spacetime where the Einstein equations break down, signaled by the divergence of physically meaningful quantities, such as curvature invariants.
To prevent such occurrences from shattering the deterministic picture of classical evolution in gravitational collapse, for which quantum physics is in principle uncalled for, Penrose put forward the weak cosmic censorship conjecture (wCCC)~\cite{Penrose:1969pc}.
This hypothesis posits that ---under mild assumptions, like physically reasonable matter and genericity of initial conditions--- a regular configuration cannot develop such singularities under gravitational collapse with the classical equations of motion, unless they are veiled by event horizons.\footnote{Singularities not hidden behind an event horizon are referred to as naked singularities.}
The wCCC is, in fact, a cornerstone of several major mathematical developments in General Relativity, such as Penrose inequalities~\cite{Penrose:1973um} and Hawking's area theorem~\cite{Hawking:1971tu}. Nevertheless, it remains a conjecture to this date.
See~\cite{Wald:1997wa} for a careful review and discussion on the subject, albeit somewhat outdated in its references. See also~\cite{Harada:2001nj, Joshi:2011rlc, Emparan:2020vyf} for more recent related reviews.

The wCCC has been a subject of intense scrutiny.
A violation of this hypothesis, by demonstrating the formation of naked singularities from generic regular initial data, would herald the breakdown of predictability within classical physics, since a satisfactory description of singularities is beyond the regime of applicability of the Einstein equations.\footnote{Singularities developing in modified theories of gravity obviously suffer from the same hindrance, but the main objective of many such theories is to avoid singularities altogether.}
The literature on the subject is vast. While a proof in favor of the wCCC is still lacking, many attempts have been made to find explicit violations, but no conclusive, physically reasonable and generic, counter-example has been found in four spacetime dimensions, so far.

It is natural to wonder what, if any, is the impact that quantum corrections on the black hole spacetime can have on the status of the wCCC. It seems particularly relevant to understand whether quantum effects strengthen or instead disfavor the censorship hypothesis. That is the subject of the present paper.
This study is made possible by a recent advancement in our understanding of quantum-corrected BHs reported in~\cite{Emparan:2020znc}, where an \textit{exact} solution of a fully quantum backreacted BH was described.
For technical reasons, such a solution is known only in three spacetime dimensions so far. Nevertheless, it offers a unique opportunity to assess the stability of quantum BHs. Here, we shall do so by testing the wCCC with a thought experiment originally envisaged by Wald, which we will now review, alongside with some of the most significant developments in this rich topic since then.

\medskip
There is a fifty-year long history of tests of the wCCC, starting with initial work by Penrose~\cite{Penrose:1973um} and quickly followed by the influential paper by Wald~\cite{Wald:1974hkz}.
Wald ventured to destroy the event horizon of an extremal Kerr-Newman BH spacetime\footnote{The Kerr-Newman family of geometries~\cite{Newman:1965my} is the most general asymptotically flat, stationary eletro-vacuum solution of Einstein-Maxwell theory in four spacetime dimensions, possessing both angular momentum and electric charge, in addition to mass~\cite{Mazur:1982db}. When the (sum of the squares of the) spin and charge are bounded from above by the (squared) mass, the spacetime features an event horizon which encloses the singularity, and such solutions are said to be underextremal, or extremal, if the bound is saturated. Otherwise, if the spin or charge are too large compared to the mass, the geometry is referred to as overextremal and it corresponds to a naked singularity.} by dropping test particles into it, in an effort to overspin, or overcharge, the BH. It was proven that such challenges to the wCCC are doomed to fail in the strict test particle limit.

Later attempts to violate the wCCC in the same vein, still adopting test particles falling into extremal BHs, considered background geometries in spacetime dimensions different from four~\cite{Bouhmadi-Lopez:2010yjy, Rocha:2011wp, Rocha:2014jma, Revelar:2017sem} and/or with the inclusion of a cosmological constant~\cite{Rocha:2011wp, Zhang:2013tba, Rocha:2014jma, Jiang:2023xca}.
The subject took an intriguing turn when counter-examples to the wCCC were proposed by starting with \textit{near-extremal} BH configurations on which test particles were made to impinge, suggesting that small regions in the space of initial data could lead to violations of the conjecture (see ~\cite{Hubeny:1998ga, deFelice:2001wj, Jacobson:2009kt, Saa:2011wq, Gao:2012ca, Duztas:2016xfg} and \cite{Revelar:2017sem}).
Some studies also proposed wCCC violation can occur by quantum tunneling~\cite{Matsas:2007bj, Richartz:2011vf}. All these hinted nonobservances of weak cosmic censorship with test particles seem to be invalidated once backreaction effects are taken into account~\cite{Hod:2002pm, Hod:2008zza, Barausse:2010ka, Barausse:2011vx, Isoyama:2011ea, Zimmerman:2012zu, Colleoni:2015afa, Colleoni:2015ena}.
This possibility was finally confirmed by Sorce and Wald~\cite{Sorce:2017dst}, who proved that the wCCC is satisfied in all attempts to destroy a BH event horizon by allowing physically reasonable matter to fall in, as long as backreaction effects (of quadratic order in the test particle parameters) are properly folded in the analysis. 

Studies of wCCC violation with test \textit{fields} ---as opposed to test \textit{particles}--- have also been conducted, with similar outcomes~\cite{Duztas:2013wua, Duztas:2016xfg, Gwak:2018akg, Chen:2018yah, Duztas:2019ick, Jiang:2023xca}. Also in this case, any counter-examples reported are dismissed if backreaction effects are accordingly incorporated~\cite{Natario:2016bay, Natario:2019iex, Goncalves:2020ccm}.
The gravitational collapse of \textit{thin matter shells} onto BHs~\cite{Mann:2008rx, Delsate:2014iia, Rocha:2017uwx} has been considered as an alternative to circumvent the limitations of the test-particle or test-field approximation, yielding exact dynamical solutions of the field equations, although representing an idealized situation in which matter is compressed on a infinitely thin surface.
The direct gravitational collapse of \textit{matter clouds}~\cite{Eardley:1978tr, Christodoulou:1984mz, Ori:1987hg, Shapiro:1991zza, Lemos:1991uz, Joshi:1993zg, Harada:1998wb} or of \textit{scalar fields}~\cite{Choptuik:1992jv, Christodoulou:1994hg} into naked singularities has also been a recurrent theme in the weak cosmic censorship challenge for the past forty-five years.\footnote{Refer to the reviews~\cite{Harada:2001nj, Joshi:2011rlc} for a more complete list of references on this topic.} However, all of these proposed violations of the wCCC involve one form or another of unstable fine-tuning (for instance, spherical symmetry or self-similarity), and some of the matter models considered in the first class of studies are regarded as physically unreasonable.

Nevertheless, in specific contexts naked singularities can in fact develop generically from regular initial data. One robust class in which this has been observed with numerical simulations involves the time evolution of black strings~\cite{Lehner:2010pn}, black rings~\cite{Figueras:2015hkb}, ultraspinning BHs~\cite{Figueras:2017zwa} or glancing BH collisions~\cite{Andrade:2018yqu, Andrade:2019edf, Andrade:2020dgc}. All these examples of cosmic censorship violation rely essentially on the Gregory-Laflamme instability~\cite{Gregory:1993vy} and, as such, only arise in spacetime dimensions larger than four.\footnote{See also Refs.~\cite{Eperon:2019viw} and~\cite{Corelli:2022phw} for recent suggestive proposals of wCCC violation.} In any case, Ref.~\cite{Emparan:2020vyf} argued that these few generic examples of mechanisms producing transient naked singularities from regular initial configurations entail a fairly mild loss of predictivity within classical physics. The only other standing claims of wCCC violation with a generic character we are aware of were developed in holographic models and are specific to asymptotically AdS spacetimes~\cite{Horowitz:2016ezu, Crisford:2017zpi}, being entirely avoided if another apparently unrelated conjecture ---the weak gravity conjecture~\cite{Arkani-Hamed:2006emk}--- holds.

\medskip
We now turn our attention to the object of our study, namely the quantum backreacted BH of Ref.~\cite{Emparan:2020znc}, on which we perform Wald's Gedanken experiment~\cite{Wald:1974hkz}. This spacetime was derived in the context of a braneworld scenario, taking the rotating AdS$_4$ C-metric as the exact bulk solution. The resulting braneworld geometry is a 3D BH that generalizes the well-known rotating BTZ spacetime~\cite{Banados:1992wn, Banados:1992gq}, and for this reason has been refered to as the qBTZ black hole. When the brane is pushed to the boundary of AdS$_4$, its tension vanishes and one recovers the BTZ metric. This is a decoupling limit, in the sense that the quantum fields do not backreact on the spacetime geometry. Otherwise, the geometry depends on a supplementary parameter, in addition to mass and angular momentum, which is associated to the backreaction of the quantum conformal fields supporting the solution. These three parameters must satisfy a certain inequality for the geometry to feature an event horizon covering the central curvature singularity. When the inequality is saturated, the corresponding BH has vanishing temperature and, similarly to the Kerr-Newman case, is said to be extremal. We will test the validity of the wCCC ---applied to the qBTZ black hole--- with inspiralling test particles \textit{on the braneworld}, in an attempt to overextremize the quantum BH and so destroy its event horizon.
Naturally, when the brane tension is taken to vanish we reproduce the results of~\cite{Rocha:2011wp} derived for the classical BTZ background.

Before moving on to the technical computations, we leave a cautionary remark: this and similar attempts to destroy BH horizons with test particles do not necessarily imply violation of the wCCC in case they are successful. Such conclusion rests on the assumption that the family of spacetimes considered, whose extremal or under-extremal representatives comprise BHs, but with over-extremal configurations yielding naked singularities, is the {\em unique} end state of the process. This was already stressed in Wald's original paper~\cite{Wald:1974hkz}. For Einstein-Maxwell theory, the only stationary vacuum BH solutions belong to the Kerr-Newman family. 
However, in our setup there is no evidence supporting the idea that the qBTZ geometry is the unique stationary fully backreacted BH solution. A process involving the capture of test particles by a qBTZ black hole and leading to a final configuration whose charges would correspond to a naked singularity might just as well signal development of an unstable configuration that might transition (by radiating away more angular momentum than mass) into a final stable and under-extremal qBTZ black hole.
One should also keep in mind that the analysis we perform here follows along the lines of~\cite{Wald:1974hkz} and therefore neglects self-force and finite size effects, as well as gravitational radiation. In the event that a parametrically small violation of the wCCC were observed, these effects should necessarily be considered.

\medskip
The rest of the paper is organized as follows. In Section~\ref{sec:qBTZ} we present the background metric we adopt to test the wCCC and discuss the conditions for the geometry to be extremal, as well as some other restrictions on its parameters. Section~\ref{sec:TestWCCC} is devoted to the actual test of the wCCC, where we provide both perturbative and numerical analyses. Conclusions and some further discussion are offered in Section~\ref{sec:Conclusion}.

\section{The backreacted (quantum) BTZ metric\label{sec:qBTZ}}

An exact construction of a BH localized on a brane was initially presented in~\cite{Emparan:1999fd}. By exploiting the AdS$_4$/CFT$_3$ duality, the authors of~\cite{Emparan:2002px} later interpreted such solution as a \emph{quantum} BTZ black hole that incorporates the backreaction of the conformal fields on the brane. A deeper analysis of the backreaction effect and the effective three-dimensional gravitational theory on the brane has been developed in~\cite{Emparan:2020znc}, together with the construction of the rotating generalization. 
These studies opened up the venue for further exploration of quantum BHs in different setups~\cite{Emparan:2022ijy, Panella:2023lsi} and their thermodynamics~\cite{Frassino:2022zaz}. 

In this paper, since we aim to push such a quantum BH beyond extremality, we will focus on the rotating metric and use the notation introduced in~\cite{Emparan:2020znc} for the rotating qBTZ solution. Moreover, since we will be working with the backreacted 3D metric, we use ${\cal G}_3$ to indicate the 3D Newton constant in the effective 3D theory. See~\cite{Bueno:2022log} for more details about the induced effective theory on the brane.

\subsection{Metric of the rotating qBTZ}

The stationary AdS$_4$ C-metric \cite{Plebanski:1976gy} from which the metric of the rotating qBTZ is obtained reads~\cite{Emparan:2020znc}
\begin{equation}
\begin{split}
    \dd s^2 = \frac{\ell^2}{(\ell + r x)^2} \biggl[&-\frac{H_{\text{C}}(r)}{\Sigma(r,x)}\left(\dd t + a x^2 \dd \phi \right)^2 + \frac{\Sigma(r, x)}{H_{\text{C}}(r)}\dd r^2 \\
    &+r^2 \left(\frac{\Sigma(r, x)}{G(x)}\dd x^2 + \frac{G(x)}{\Sigma(r, x)}\left(\dd \phi - \frac{a}{r^2} \dd t\right)^2 \right)\biggr] \, ,
\end{split}
\label{AdS4Cmetricgeneral}
\end{equation}
where $\ell^{-1}$ is the inverse of the tension of the brane, which is localized at $x = 0$, while $a$ is related to the BH spin. The metric functions are 
\begin{subequations}
\begin{align}
    &H_{\text{C}}(r) = r^2 \left(\frac{1}{\ell_4^2}-\frac{1}{\ell^2} \right)+\kappa -\frac{\mu \ell}{r}+\frac{a^2}{r^2}\,, \label{HfunctionCmetric4D}\\
    &G(x) = 1-\kappa x^2 -\mu x^3 + a^2 \left(\frac{1}{\ell_4^2}-\frac{1}{\ell^2} \right) x^4\,, \label{Gxgeneral}\\
    &\Sigma(r, x)= 1+\frac{a^2x^2}{r^2}\,.
\end{align}
\end{subequations}
$\kappa$ is a discrete parameter taking the values $\kappa =-1,\, 0,\, 1$, while $\mu$ is classically related to the BH mass in four dimensions. Additionally, $G(x)$ being a fourth-order polynomial, we will consider, as usual, a configuration such that it has at least one positive root, the smallest of which is defined as $x_1$. 

By projecting Eq.~\eqref{AdS4Cmetricgeneral} onto the brane at $x = 0$, one obtains the metric of the rotating qBTZ (see Ref.~\cite{Emparan:2020znc} for details on the subtleties involved in this procedure)
\begin{equation}\label{metricgeneral}
\dd s^2 = g_{tt} \, \dd t^2 + g_{\phi\phi} \, \dd \phi^2 + 2 g_{t\phi} \, \dd t \, \dd \phi + g_{\hat{r}\hat{r}} \, \dd \hat{r}^2\, ,
\end{equation}
where $t$ and $\phi$ correspond to  the physical coordinates $\bar t$ and $\bar \phi$ in~\cite{Emparan:2020znc}, while $\hat{r}$ corresponds to the original $r$ coordinate in~\cite{Emparan:2020znc}. The metric functions are explicitly given by
\begin{subequations}
\begin{align}
& g_{tt} = -\left(\frac{r^2}{\ell_3^2}-8{\cal G}_3 M -\frac{\mu \ell \Delta^2}{\hat{r}} \right)\,,\\
& g_{\phi\phi} = r^2 + \ell_3^2 \, \frac{\mu \ell \tilde a^2 \Delta^2}{\hat{r}}\,,\\
& g_{t\phi} = -4{\cal G}_3 J \left(1+\frac{\ell}{\hat{r} x_1} \right)\,, \\
& g_{\hat{r}\hat{r}} = \frac{1}{H(r)} = \left[\frac{r^2}{\ell_3^2}-8{\cal G}_3 M + \frac{(4{\cal G}_3 J)^2}{r^2} -\mu \ell (1-\tilde a^2)^2\, \Delta^4 \, \frac{\hat{r}}{r^2}\right]^{-1}\,,\label{grrgeneral}
\end{align}
\end{subequations}
where $\ell_3$ is the induced $\text{AdS}_3$ length on the brane, related to the $\text{AdS}_4$ length scale and the brane tension by the holographic relation $\ell_3^{-2} = \ell_4^{-2} - \ell^{-2}$.\footnote{Consistency of this relation demands that $\ell_4^{-2}>\ell^{-2}$. If one interprets $\ell^{-1}$ as the acceleration parameter of the AdS C-metric, this is the so-called slow-acceleration limit~\cite{Podolsky:2002nk}, where the acceleration horizon is absent.}
The physical meaning of the other parameters is inherited from the four-dimensional solution and will be explained below. 

First, we note that, in the three-dimensional metric, $\ell$ controls the strength of the backreaction of quantum corrections. Indeed, in the decoupling limit $\ell \rightarrow 0$, we recover the classical BTZ metric. Such a limit corresponds to pushing the brane to the asymptotic AdS boundary in the classical four-dimensional bulk. Note, however, that one actually recovers the rotating BTZ solution only in the case $\kappa = -1$. 

The line element written above, Eq.~\eqref{metricgeneral}, still mixes the physical $r$ and the unphysical $\hat{r}$ radial coordinates. From now on, we will use exclusively the physical $r$ coordinate (see \cite{Emparan:2020znc} for further details about the different coordinates), defined in terms of $\hat{r}$ as
\be
r^2 = (1-\tilde a^2) \Delta^2 \hat r^2 + r_S^2\,,
\label{rtorbar}
\ee
where $\ta$ is the same combination of parameters introduced in \cite{Emparan:2020znc} 
\begin{equation}
    \ta^2 \equiv a^2 x_1^4 \left(\frac{1}{\ell_4^2}-\frac{1}{\ell^2} \right) = \frac{a^2 x_1^4}{\ell_3^2}\, ,
    \label{tildeaCmetric}
\end{equation}
while $\Delta$ and $r_{S}$ are functions of $x_1$ and $\tilde{a}$, defined as
\begin{subequations}
\begin{align}
\Delta &= \frac{2x_1}{3-\kappa x_1^2-\tilde a^2}\,, \label{Deltaqbtz}\\
r_S &= \ell_3 \frac{2\tilde a \sqrt{2-\kappa x_1^2}}{3-\kappa x_1^2-\tilde a^2}\,. 
\end{align}
\label{Delta_rS}
\end{subequations} 
\!\!The only other parameter appearing in~\eqref{metricgeneral} is $\mu$. From the three-dimensional braneworld perspective it parametrizes the corrections due to the backreaction of quantum conformal fields. As in Ref.~\cite{Emparan:2020znc}, $\mu$ implicitly defines the smallest positive root $x_1$ of Eq.~\eqref{Gxgeneral}:
\be
\mu = \frac{1-\kappa x_1^2+\tilde a^2}{x_1^3}\,.
\label{mugeneralqBTZ}
\ee
Finally, the mass $M$ and the angular momentum $J$ of the qBTZ solution read: 
\begin{subequations}
\begin{align}
M &= \frac{1}{2{\cal G}_3}\frac{-\kappa x_1^2+\tilde a^2(4-\kappa x_1^2)}{(3-\kappa x_1^2-\tilde a^2)^2}\,, \label{M}\\
J &= \frac{\ell_3}{{\cal G}_3}\frac{\tilde a(1-\kappa x_1^2+\tilde a^2)}{(3-\kappa x_1^2-\tilde a^2)^2}\,. 
\label{J}
\end{align}
\label{parameters}
\end{subequations}

In the following, we will work with the metric functions, expressed in terms of the parameters $x_1$ and $\tilde a$ (see \cref{App:MetricFunctionsqBTZ}). We will restrict to $\ta <1$ in order to avoid naked closed timelike curves (see Ref.~\cite{Emparan:2020znc}). We will consider both the $\kappa = \pm 1$ cases (the case $\kappa = 0$ is not considered separately, as it is automatically recovered by taking the limit $x_1 \to 0$) and, without loss of generality, we will restrict to $J>0$ configurations.

\subsection{Horizons and extremality condition}
\label{sec:horizonsandextremality}

As described in the introduction, our aim is to assess whether a test particle falling into an extremal rotating qBTZ black hole can destroy its event horizon, thereby exposing the curvature singularity within. This section, therefore, is dedicated to the study of the conditions that define extremal solutions.

\begin{figure}[t]
\centering
\subfigure[$\kappa =-1$]{\includegraphics[width=7.5cm]{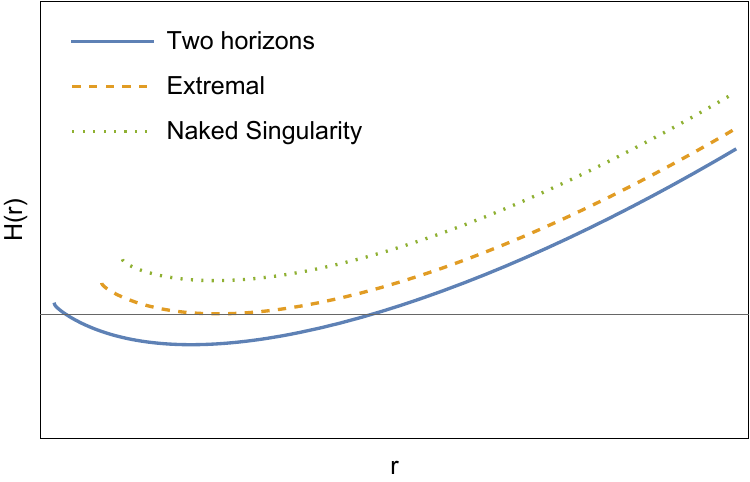}}
\hfill
\subfigure[$\kappa = 1$]{\includegraphics[width=7.5cm]{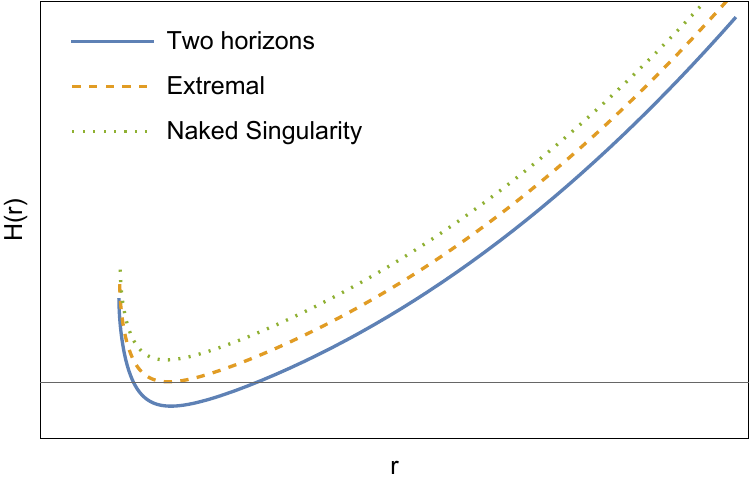}}
\caption{Qualitative behavior of the metric function $H(r)$ for $\kappa = \pm 1$. Depending on the values of the parameters, one obtains a non-degenerate horizon (solid blue lines), an extremal horizon (dashed orange lines) or a naked-singularity (dotted green lines). }
\label{Hqualitative}
\end{figure}

As usual, the radial location of the horizon is given by the largest root of the metric component $g^{\hat{r} \hat{r}} = H(r)$. Inspection of Eq.~\eqref{grrgeneral} shows that $H(r)$ has a minimum in the positive $r$-axis.
Additionally, due to the coordinate redefinition \eqref{rtorbar}, $\hat{r} = 0$ gets mapped to $ r = r_S$ and so we have to cut the $r$-axis at $r_S$. The radius at which $H$ attains its minimum, $r_\text{min}$, is anyway always in the interval $ r \in [r_S, \infty)$. The presence or absence of horizons depends, therefore, on the sign of $H(r_\text{min})$:
\begin{itemize}
\item If $H(r_\text{min})<0$, the metric possesses a non-degenerate horizon;
\item If $H(r_\text{min})=0$, the horizon is degenerate and the BH is extremal;
\item If $H(r_\text{min})>0$, horizons are absent and the spacetime features a naked singularity. 
\end{itemize}
Qualitative plots of the behavior of $H(r)$ are shown in Fig.~\ref{Hqualitative} for $\kappa = \pm 1$. 

The extremality condition is obtained when the horizon coincides with the minimum of the metric function, namely when the conditions $H(\rH) =0$ and  $H'(\rH) = 0$ are satisfied simultaneously (prime will refer to derivatives with respect to $r$). We will encompass both the $\kappa = 1$ and $\kappa = -1$ cases in our analysis, and it is convenient to work with the function $Q(r) \equiv r^2 H(r)$:
\begin{equation}
Q(r) = \frac{r^4}{\ell_3^2}-8 {\cal G}_3 M r^2 + (4 {\cal G}_3 J)^2 -\mu \ell (1-\tilde a^2)^{3/2} \Delta^3 \sqrt{r^2-r_S^2}\, .
\end{equation}
The extremality condition, then, corresponds to requiring $Q(\rH) = Q'(\rH) = 0$. This yields the following system of equations:
\begin{subequations}
\begin{align}
Q(\rH) &= \frac{\rH^4}{\ell_3^2}-8 {\cal G}_3 M \rH^2 + (4 {\cal G}_3 J)^2 -\mu \ell (1-\tilde a^2)^{3/2} \Delta^3 \sqrt{\rH^2-r_S^2} = 0\,,\\
Q'(\rH) &= \frac{4\rH^2}{\ell_3^2}-16 {\cal G}_3 M -\frac{\mu \ell \, (1-\tilde a^2)^{3/2}\Delta^3}{\sqrt{\rH^2-r_S^2}} = 0\, .
\end{align}
\end{subequations}

Solving the above for $\rH$ and $\mu$, and using Eq.~\eqref{parameters} to express the results as functions of the parameters $\ta$ and $x_1$, yields
\begin{subequations}
\begin{align}
&\rH^2 = \frac{2}{3} \ell_3^2 \left[\frac{(1-\ta^2)\sqrt{12\ta^2 + x_1^4}}{\left(3-\kappa x_1^2 -\ta^2 \right)^2}-\frac{x_1^2 \left(5 \ta^2 \kappa +\kappa \right)-12 \ta^2}{\left(3-\kappa x_1^2 -\ta^2\right)^2}\right]\,,
\label{rHextremalkappa}\\
&\mu \nu= \frac{1}{3}\sqrt{\frac{2}{3}} \left(\sqrt{\frac{12 \ta^2}{x_1^4}+1}+2 \kappa \right) \sqrt{\sqrt{\frac{12 \ta^2}{x_1^4}+1}-\kappa }\, , \label{muextremalkappa}
\end{align}
\end{subequations}
where we introduced the rescaled backreaction parameter,
\begin{equation}
\nu \equiv \ell/\ell_3\,.
\end{equation}
Clearly, the zero-backreaction limit $\ell\to0$ corresponds to $\nu \to 0$.

\begin{figure}[!h]
\centering
\includegraphics[width=0.6\textwidth]{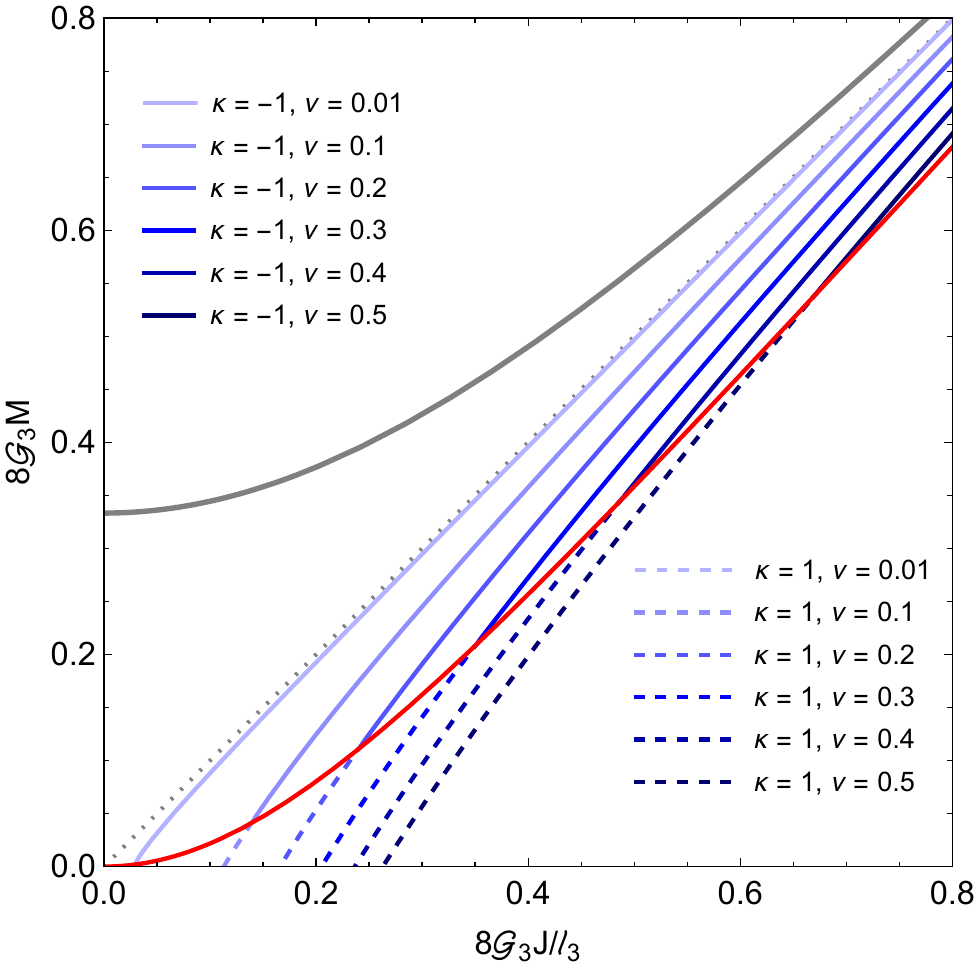}
\caption{$M$ vs $J$ diagram. The gray solid line represents the upper bound on the masses of the qBTZ configurations. The dotted gray diagonal represents the classical BTZ extremality condition $M = J/\ell_3$. The solid red line corresponds to the $\kappa = 0$ line, separating solutions with $\kappa = 1$ (below) from those with $\kappa = -1$ (above). The blue lines denote extremal solutions: solid lines correspond to extremal configurations with $\kappa = -1$, while dashed lines to $\kappa = 1$ ones. The dashed line for the $\nu = 0.01$ case is shown, but not visible, since it is very close to the horizontal axis.}
\label{DiagramMvsJ}
\end{figure}

Having the two expressions~\eqref{rHextremalkappa} and~\eqref{muextremalkappa}, we can now plot the corresponding extremality condition in the $(M,J)$ parameter space (see Fig.~\ref{DiagramMvsJ}). Configurations falling on the left side of the blue lines are non-degenerate BHs, while those living on the right side correspond to naked singularities.
The diagram $M$ vs $J$ in Ref.~\cite{Emparan:2020znc} reported only the classical BTZ extremality line and the curve corresponding to the upper bound on the mass implied by the holographic construction of the qBTZ solution (grey dotted and solid lines in Fig.~\ref{DiagramMvsJ}, respectively). Here, we have complemented this diagram by also adding the curve (red solid line in Fig.~\ref{DiagramMvsJ}) along which the two families of solutions with $\kappa = -1$ and $\kappa = 1$ smoothly join. This curve is obtained by setting $\kappa = 0$ in Eqs.~\eqref{M} and~\eqref{J}.

A good check on the above results is to study their vanishing backreaction limit $\nu \to 0$, for which the qBTZ extremality conditions should reduce to the classical BTZ ones, $J = M \ell_3 $ and $\rH^2 =4 \ell_3^2 {\cal G}_3 M$ (see \cref{App:ClassicalBTZ}). These limits are realized very differently for $\kappa = 1$ and $\kappa = -1$. 

For $\kappa=1$, we must require $\ta = 0$, and then Eq.~\eqref{rHextremalkappa} reduces to zero. This can be understood as the fact that $M$ given by Eq.~\eqref{rHextremalkappa} can only be positive\footnote{If we restrict to positive angular momentum, the classical extremality condition implies that also $M$ must be positive.} when $x_1 = 0$, so $\rH^2 =4 \ell_3^2{\cal G}_3 M$ is still satisfied, but with $M = 0$. We also note that, in the classical case, $M = 0$ represents the transition from positive-mass BHs to negative-mass naked singularities. 

For $\kappa = -1$, we must require $x_1 = \sqrt{2\ta}$. Indeed, we straightforwardly get $\rH^2 =4 \ell_3^2{\cal G}_3 M$ from Eq.~\eqref{rHextremalkappa}, using Eq.~\eqref{M}.

The limit of zero backreaction can also be understood visually, with the help of Fig.~\ref{DiagramMvsJ}. The blue lines (quantum extremal configurations) get closer and closer to the dotted diagonal line (the classical extremality condition) when $\nu$ approaches zero.
For $\kappa=0$, the $\nu\to0$ limit is realized by sliding the point on the red curve where the two extremal lines (for $\kappa = \pm 1$) join, in the direction of decreasing $\nu$. It is then apparent that the no-backreaction limit is attained when the dashed blue lines (for $\kappa=+1$) collapse at $J = 0$ and $M = 0$, which in turn implies $\ta = 0$ and $x_1 = 0$.

\subsection{Further constraints on the parameters}
\label{sec:constraintsparameters}

An important ingredient before testing wCCC is to understand and analyze in detail the parameter space related to the configurations of interest. In the following, we will provide additional constraints on the parameters $\ta$ and $x_1$, which supplement those analyzed in~\cite{Emparan:2020znc}.

On the $\kappa=-1$ branch, Eq.~\eqref{mugeneralqBTZ} reduces to $\mu = x_1^{-3}(1+\ta^2+x_1^2)$, which is a manifestly positive quantity. In order for the extremality condition~\eqref{muextremalkappa} to hold, we must impose
\begin{equation}
\sqrt{\frac{12 \ta^2}{x_1^4}+1}-2 > 0 \, \qquad \Rightarrow \qquad x_1 < \sqrt{2\ta}\, .
\end{equation}
Therefore, the parameter intervals of interest are $0\leq x_1< \sqrt{2\ta}$ and $0<\ta<1$, as already mentioned below Eq.~\eqref{parameters}. In the $\kappa = -1$ case, these constraints also cast an upper bound on the backreaction parameter $\nu$, which can be found by inverting Eq.~\eqref{muextremalkappa}:
\begin{equation}
    \nu = \left[\frac{1}{3}\sqrt{\frac{2}{3}} \left(\sqrt{\frac{12 \ta^2}{x_1^4}+1}-2 \right) \sqrt{\sqrt{\frac{12 \ta^2}{x_1^4}+1}+1 }\,\right]\, \frac{x_1^3}{1-\kappa x_1^2 + \ta^2}\, .
    \label{nuextremalityqBTZ}
\end{equation}
This function is monotonically decreasing in $x_1$, so its maximum at fixed $\ta$ is attained at $x_1 = 0$, from which we get $\nu = 4 \ta^{3/2}/\left[3^{3/4} (1+\ta^2)\right]$. The latter is, instead, a monotonically increasing function in $\ta$, whose maximum occurs at $\ta = 1$, from which we get $\nu < 2/3^{3/4}\simeq 0.8774$.

In the $\kappa = 1$ case, as long as we restrict to positive values of the angular momentum $J$ \eqref{J}, we have again an upper bound on $x_1$. Indeed $1-x_1^2+\ta^2\geq0$ implies $0\leq x_1 \leq \sqrt{1+\ta^2}$. Then $\mu>0$ and the extremality condition \eqref{muextremalkappa} has automatically the right sign on both sides. As for $\nu$, it is not upper bounded anymore. For large values of $\nu$, i.e., $\nu \to \infty$, the extremality condition \eqref{muextremalkappa} reduces to $1-x_1^2+\ta^2 = 0$, namely to the line $x_1 = \sqrt{1+\ta^2}$. \\

Finally, while for $\kappa = -1$ the mass \eqref{M} is always positive, for $\kappa = 1$ the mass $M$ can be either positive or negative. In the classical BTZ geometry, negative mass states (above a certain minimum to be discussed below) correspond to naked singularities. Here, instead, the quantum effects are able to generate horizons despite $M$ being negative. From Eq.~\eqref{M}, negative mass states are obtained when $x_1^2 > 4\ta^2/(1+\ta^2)$. Therefore, we have
\begin{equation}
\begin{split}
0\leq \, &x_1^2 < \frac{4\ta^2}{1+\ta^2} \qquad \qquad\,  \text{for positive}\, \, M\,, \\
\frac{4\ta^2}{1+\ta^2} < \, &x_1^2 \leq 1+\ta^2 \qquad \qquad \, \, \text{for negative}\, \, M\,.
\end{split}
\end{equation}
If $x_1^2 = 4\ta^2/(1+\ta^2)$, $M=0$, but $J$ is not zero. This configuration can also have two horizons and a related extremal configuration. This is due entirely to the quantum corrections, since a classical BTZ configuration with $M=0$ corresponds to a naked singularity. 

There is another particular case, represented by the state with the \emph{least} mass. This is obtained with $x_1^2 = 1+\ta^2$, which gives $M = -1/(8\mathcal{G}_3)$. At extremality, from Eq.~\eqref{muextremalkappa}, we get $\ta = 0$ (and also $J = 0$). Therefore, the horizon radius of the extremal configuration \eqref{rHextremalkappa} is zero. This is analogous to the $M = 0$ classical state.

\section{Testing Weak Cosmic Censorship\label{sec:TestWCCC}}

\subsection{Particle motion in the qBTZ metric and maximum angular momentum}
\label{sec:GeodesicsqBTZ}

To test the wCCC, we will consider particles on geodesics falling into the extremal, rotating qBTZ black hole. However, those with sufficiently large angular momentum will not reach the BH event horizon due to the high centrifugal barrier. Therefore, it is crucial to determine the maximum value of the angular momentum allowing for particle capture. 

To that end, consider the Lagrangian of a point-like particle
\begin{equation}\label{Lagrangianparticle}
{\mathscr{L}} = \frac{1}{2}g_{\mu\nu} \dot x^\mu \dot x^\nu\, = \frac{1}{2}\left[g_{tt} \, \dot t^2 + g_{\phi\phi} \, \dot \phi^2 + 2 g_{t\phi}\, \dot t \dot \phi + g_{rr} \, \dot r^2 \right].
\end{equation}
Invariance under time translation and rotation of the metric \eqref{metricgeneral} implies the existence of two conserved quantities, associated with the conjugate momenta
\begin{subequations}
\begin{align}
&p_t = \partial_{\dot t} \mathscr{L} = g_{tt} \dot t + g_{t\phi} \dot \phi = -E\,, \\
&p_\phi =\partial_{\dot \phi} \mathscr{L} = g_{\phi\phi} \dot \phi + g_{t\phi} \dot t = L\,.
\end{align}
\label{conjugatemomenta}
\end{subequations}
$E$ can be interpreted as the energy of the particle, while $L$ represents its angular momentum (per unit mass). 
Inverting the above relations gives
\begin{subequations}
\begin{align}
&\dot t = \frac{g_{t\phi}L+g_{\phi\phi}E}{g_{t\phi}^2-g_{tt}g_{\phi\phi}}\,, \label{tdot}\\
&\dot \phi = -\frac{g_{tt}L+g_{t\phi}E}{g_{t\phi}^2-g_{tt}g_{\phi\phi}}\, . \label{phidot}
\end{align}
\label{dottdotphi}
\end{subequations}

To compute the maximum value of $L$ allowing particles to be absorbed by the BH, we follow Wald \cite{Wald:1974hkz} and simply impose the geodesics to be future directed, namely $\dot t>0$, which implies, from Eq.~\eqref{tdot},
\begin{equation}
\begin{split}\label{Lbound}
&g_{t\phi}L+g_{\phi\phi}E > 0
\qquad \Rightarrow \qquad
L < -\frac{g_{\phi\phi}}{g_{t\phi}}E\, .
\end{split}
\end{equation}

If we want the particle to be captured by the BH, this has to hold true everywhere outside the event horizon, $r = \rH$. Precisely at the event horizon the upper bound on the particle's angular momentum becomes
\begin{equation}
L < \frac{\left(-\tilde a^2-\kappa x_1^2+3\right)^2 \left(\rH^2+\frac{8 \tilde a^2 \sqrt{1-\tilde a^2} \nu  \ell _3^3 \left(1+\tilde a^2-\kappa x_1^2\right)}{\left(-\tilde a^2- \kappa x_1^2+3\right)^2 \sqrt{\left(-\tilde a^2- \kappa x_1^2+3\right)^2 \rH^2+4 \tilde a^2 \left(\kappa x_1^2-2\right) \ell _3^2}}\right)}{4 \tilde a \ell _3 \left(1+\tilde a^2-\kappa x_1^2\right) \left(\frac{2 \sqrt{1-\tilde a^2} \nu  \ell _3}{\sqrt{\left(-\tilde a^2-\kappa x_1^2+3\right)^2 \rH^2+4 \tilde a^2 \left(-2+\kappa x_1^2\right) \ell _3^2}}+1\right)}\, E\, .
\label{upperboundL}
\end{equation}
If this condition is not met the particles will fail to be captured by the BH. 

The inequality~\eqref{upperboundL} depends both on the particle energy $E$ and on background quantities $(\tilde a, x_1, \nu, \ell_3, \kappa)$. 
Since we are interested in testing whether the BH can be overspun, we will restrict our considerations to particles infalling in an extremal BH background. Replacing our expression for the extremal event horizon~\eqref{rHextremalkappa} in Eq.~\eqref{upperboundL} yields
\begin{equation}
L_\text{max}=\frac{4 \ell _3 \left[\frac{\sqrt{6} \ta^2 \nu  \left(\ta^2-\kappa  x_1^2+1\right)}{\sqrt{\sqrt{12 \ta^2+ x_1^4}-\kappa  x_1^2}}+\frac{1}{6} \left(\left(1-\ta^2\right) \sqrt{12 \ta^2+x_1^4}-\left(5 \ta^2+1\right) \kappa  x_1^2+12 \ta^2\right)\right]}{\ta \left(\ta^2-\kappa  x_1^2+1\right) \left(\frac{\sqrt{6} \nu }{\sqrt{\sqrt{12 \ta^2+ x_1^4}-\kappa  x_1^2}}+1\right)}\, E\, .
\label{Lmax}
\end{equation}
$L_\text{max}$ is positive for both $\kappa = \pm 1$ in the parameter ranges considered.

\subsection{The effect of particle capture on extremal qBTZ black holes}

In what follows, we allow an extremal BH to absorb a particle with energy $E$ and angular momentum \eqref{Lmax}. We then study what configuration the system can settle down to, after the particle is captured. Assuming the end state is also a stationary spacetime described by the same class of qBTZ geometries\footnote{These assumptions require dedicated analyses for validation, but suh a study is beyond the scope of the present paper. Contrary to the Kerr-Newman case, there is no guarantee of uniqueness for these quantum BHs, and our only supporting argument at the moment is that the qBTZ spacetime is the only known stationary BH in 3D that exactly incorporates backreaction from quantum conformal fields.}, there are only three options available: either a non-degenerate BH, an extremal BH, or a horizonless (naked singularity) configuration. 

At extremality, the minimum of the metric function $H(r)$ coincides with the horizon radius. After the absorption of a particle, the minimum will shift and may not correspond to an event horizon anymore. A simple way to test the wCCC is to evaluate the sign of $H(r)$ computed at its minimum $r_\text{min}$ after absorption:
\begin{itemize}
\item If $H(r_\text{min}) < 0$, one obtains a non-degenerate horizon and the singularity is shielded;
\item If $H(r_\text{min}) = 0$, the black hole remains extremal;
\item If $H(r_\text{min})>0$, the extremal BH transitions to a naked singularity.
\end{itemize}

\subsubsection{Perturbative approach}
\label{subsec:PerturbativeApproachqBTZ}

In this section, we employ a perturbative approach and focus on analyzing the linear response of the extremal background after the absorption of a test particle with maximum angular momentum. 

At linear order, after a small perturbation the BH parameters change as $\ta \to \ta + \delta \ta$ and $x_1 \to x_1 + \delta x_1$, implying also a shift on the location of the minimum of function $H$, $r_\text{min} \to r_\text{min} + \delta r$. This metric function, evaluated at its minimum, changes accordingly,
\begin{equation}\label{Hexpansiongeneral}
H(r_\text{min}+\delta r, \ta + \delta \ta, x_1 + \delta x_1) = H(r_\text{min}, \ta, x_1) + \delta H\, ,
\end{equation}
where $\delta H$ reads as 
\begin{equation}\label{deltaH}
\delta H = \frac{\partial H}{\partial r}\biggr|_{r=r_\text{min}} \!\! \delta r \; + \;  \frac{\partial H}{\partial {\ta}}\biggr|_{r=r_\text{min}} \!\! \delta \ta \; + \; \frac{\partial H}{\partial {x_1}}\biggr|_{r=r_\text{min}} \!\! \delta x_1 \, ,
\end{equation}
and all derivatives are also understood to be computed using the values of $\tilde a$ and $x_1$ of the extremal background. By definition, the first term in $\delta H$ is zero.

At linear order, it is straightforward to express $\delta \tilde a$ and $\delta x_1$ in terms of the particle energy and angular momentum. Indeed, after the absorption of the particle, the BH mass $M$ and angular momentum $J$ change as
\begin{equation}
\begin{split}
&M \to M + E \equiv M + \delta M\, ,\\
&J \to J + L_\text{max} \equiv J + \delta J\, .
\end{split}
\label{MJchange}
\end{equation}
The quantities $\delta M$ and $\delta J$ depend on $\delta \ta$ and $\delta x_1$ through the variation of Eqs.~\eqref{M} and~\eqref{J}, 
\begin{subequations}
\begin{align}
\delta M &= \partial_{\ta} M \, \delta \ta + \partial_{x_1} M \, \delta x_1\, ,\\
\delta J &=  \partial_{\ta} J \, \delta \ta + \partial_{x_1} J \, \delta x_1\, .
\end{align}
\label{deltaMdeltaJ}
\end{subequations}
It is straightforward, therefore, to invert the above and write $\delta \tilde a$ and $\delta x_1$ as functions of $\delta M$ and $\delta J$. 
Then we substitute $\delta J$ using Eq.~\eqref{Lmax}, compute $\delta H$ and study its sign. Since we consider the extremal configuration as the (initial) background, we also replace $\nu$ by the expression given in Eq.~\eqref{muextremalkappa}. 

After some lengthy but straigthforward simplifications, one obtains $\delta H = 0$ for \emph{any} value of $\nu$ (a more detailed explanation of this result is reported in \cref{App:deltaHzero}) and independently of $\kappa = \pm 1$.\footnote{In the classical BTZ spacetime ($\nu = 0$), a linear analysis similar to ours, involving the scattering of scalar fields and yielding $\delta H = 0$ was performed in~\cite{Chen:2018yah}.} Therefore, at linear order in perturbations, an extremal BH remains extremal after absorbing test particles with the largest possible ratio $L/E$ that can be captured. Particles with $L<L_\text{max}$ impart less angular momentum to the BH and therefore will give rise to under-extremal configurations.
The singularity is therefore shielded and weak cosmic censorship respected.

\subsubsection{Numerical assessment}

We now go beyond the linear approximation and perform a numerical test of the wCCC. To be specific, we shall still consider the test particle regime, without it backreacting on the spacetime apart from shifting the mass and angular momentum of the initially extremal BH. However, and in contrast with the previous subsection, we will now consider particles of \textit{finite}, albeit small, mass compared to the BH they fall into.  

Contrary to the previous calculation, here we leave $\nu$ as a free parameter, in order to study different backreaction regimes and their impact on the wCCC. The parameter $x_1$ is then constrained by the extremality condition~\eqref{muextremalkappa} to be a certain function of $\ta$. This is done by discretizing the $\ta$-axis and numerically solving Eq.~\eqref{muextremalkappa} for each point on the chosen $\ta$-interval. This yields two arrays of $\ta$ values and their corresponding $x_1$ values at extremality. One can then use these pairs and Eq.~\eqref{M} and~\eqref{J} to compute the mass and angular momentum of the corresponding extremal configurations.
For such BH, a particle with energy $\delta M = M/100$ (with $M$ the mass of each extremal configuration) and angular momentum~\eqref{Lmax} (evaluated on the corresponding extremal configuration) is formally added to the background. The mass and angular momentum of the new configuration are computed according to Eq.~\eqref{MJchange}. Having these values, the code then solves Eq.~\eqref{M} and~\eqref{J} for the new values of $\ta$ and $x_1$ after absorption. Finally, the minimum of the metric function $H$, expressed in terms of the new values of the parameters, is found numerically and then $H(r_\text{min})$ is evaluated. 

\begin{figure}[t]
\centering
\includegraphics[width=12cm]{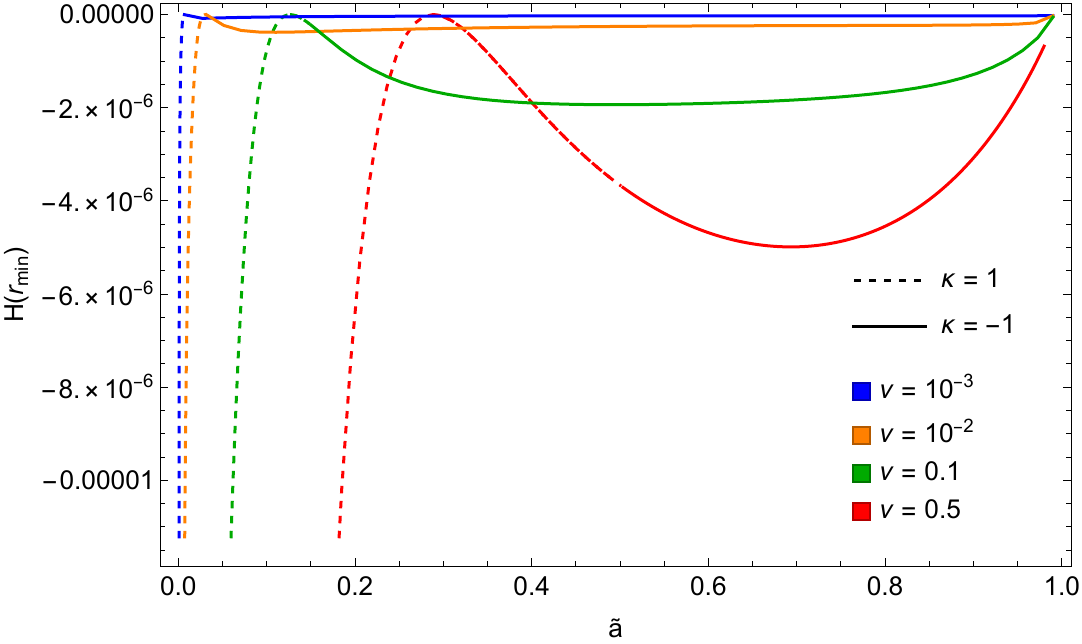}
\caption{Values of $H(r_\text{min})$ in different backreaction regimes, for configurations with $\kappa = +1$ (dashed lines) and $\kappa = -1$ (solid lines). Both positive and negative masses are considered. All the curves are below zero. The curves shown do not extend quite up to $\ta=1$ for a technical reason. If the value of $\ta$ \emph{before} the particle is absorbed is extremely close to $1$, the final value of $\ta$ after absorption will be greater than $1$, while our code is restricted to $\ta < 1$. One can extend the curves further by considering particles of smaller mass.}
\label{PositionMinimum}
\end{figure}

The results are reported in Fig.~\ref{PositionMinimum}, for different values of $\nu$, and they endorse the wCCC: the metric function at the minimum, after absorption, is negative for all values of $\nu$ considered, so a pair of horizons always form after absorption, the outer one being the event horizon, while the inner one becomes a Cauchy horizon. 

We highlight that, especially for the $\kappa = -1$ branch, which corresponds to the larger values of $\ta$ (see Fig.~\ref{DiagramMvsJ}), the stronger the backreaction effects (larger $\nu$) the \emph{more negative} $H(r_\text{min})$ becomes, indicating that large backreaction effects typically strengthen cosmic censorship.

The local maximum observed for all curves of $H(r_\text{min})$ in Fig.~\ref{PositionMinimum} is intriguing, given that it suggests that near-extremality can be preserved for specific combinations of $\ta$ and $\nu$. We do not have an intuitive explanation for this.

\section{Conclusions and outlook\label{sec:Conclusion}}

In this paper we applied Wald's well-known Gedanken experiment to the recently described qBTZ black hole, a three-dimensional spacetime that incorporates the effects of strongly-coupled quantum fields in an exact manner. Accordingly, we considered dropping test particles into an extremal BH belonging to that family of solutions, in an attempt to disrupt its event horizon and therefore expose the singularity contained within.
Surprisingly, the inclusion of the quantum backreaction of the geometry appears to have minimal influence on the outcome of this test of the wCCC: throwing a test particle at an extremal qBTZ black hole can, at most, leave it extremal, just like what happens with its classical counterpart. Nevertheless, we also showed evidence that, when going beyond the test particle limit, it is harder to violate the wCCC with the quantum corrected BH spacetime, when compared with the classical BTZ black hole. 
One might heuristically summarize this idea by saying that quantum backreaction strengthens weak cosmic censorship.

In the process of assessing the resilience of the event horizon of quantum black holes, having the weak cosmic censorship conjecture in mind, we had to perform a detailed analysis of the parameter space  of rotating qBTZ solutions, complementing that of Ref.~\cite{Emparan:2020znc}. This was instrumental in our investigations of the effect of backreaction from strongly-coupled quantum fields on the wCCC.

The thought experiment conducted in this paper has been described from the 3D point of view, in which the BH tested incorporates quantum corrections. Nevertheless, it has a holographic description, and from the 4D perspective, it corresponds to sending test particles at a classical BH in the bulk, but the particles are restricted to move on a codimension-1 braneworld. A different test of the wCCC involves particles falling more generally in the 4D bulk into the same 4D black hole, which can then be interpreted holographically from the 3D braneworld perspective. This analysis is currently in progress~\cite{FrassinoPrep}.

Concerning the three-dimensional spacetimes we have considered, expanding our study to encompass the near-extremal case would be an interesting extension of the present work. This analysis may either unveil parameter regions where overspinning is feasible or a mitigation of this behavior due to the quantum backreaction.

Furthermore, it would also be interesting to apply our analysis to the three-dimensional Kerr-de Sitter BH, recently constructed in~\cite{Panella:2023lsi} in a similar fashion to the qBTZ, using braneworld holography, as well as to the charged case~\cite{Feng:2024uia, Climent:2024nuj}. Such analyses will shed more light on the impact of quantum corrections on the wCCC. 

A natural and important question that deserves a dedicated study is whether the qBTZ spacetime is classically stable in the sense of its quasinormal spectrum only featuring decaying modes. If so, this would lend support to the idea that, even if perturbed, a qBTZ black hole should settle down to another geometry within the same qBTZ family.

Finally, another development of obvious interest would be the derivation of the five-dimensional version of the AdS C-metric. If achievable, that would presumably allow the derivation, following the rationale of~\cite{Emparan:2020znc, Emparan:2002px}, of a four-dimensional quantum corrected black hole.

\section*{Acknowledgements}
We are indebted to Roberto Emparan for interesting and fruitful discussions.
APS would also like to thank Mariano Cadoni and Selene Matta for useful discussions.
JVR is grateful for the hospitality at the Institute of Cosmos Sciences at the University of Barcelona, where part of this work was done.
JVR acknowledges financial support from \textit{Funda\c{c}\~ao para a Ci\^encia e a Tecnologia} (FCT), Portugal, under the project “MECOMOG–Mergers of compact objects in modified gravity” with reference no. 2022.08368.PTDC.

\bigskip

\begin{appendix}

\section{Explicit expressions of the qBTZ metric functions}
\label{App:MetricFunctionsqBTZ}

The metric components of the qBTZ spacetime~\eqref{metricgeneral}, expressed in terms of the parameters $x_1$ and $\tilde a$, and using the radial coordinate $r$ instead of $\hat{r}$, are the following: 
\begin{equation}
g_{tt} =-\frac{8 \sqrt{1-\ta^2} \, \nu  \, \ell _3 \left(\ta^2-\kappa  x_1^2+1\right)}{\left(\ta^2+\kappa  x_1^2-3\right)^3 \sqrt{\frac{4 \ta^2 \ell _3^2 \left(\kappa  x_1^2-2\right)}{\left(\ta^2+\kappa  x_1^2-3\right)^2}+r^2}}+\frac{16 \ta^2-4 \left(\ta^2+1\right) \kappa  x_1^2}{\left(\ta^2+\kappa  x_1^2-3\right)^2}-\frac{r^2}{\ell _3^2}\, ,
\end{equation}
\begin{equation}
g_{\phi\phi}= r^2-\frac{8 \ta^2 \sqrt{1-\ta^2} \nu  \ell _3^3 \left(\ta^2-\kappa  x_1^2+1\right)}{\left(\ta^2+\kappa  x_1^2-3\right)^3 \sqrt{\frac{4 \ta^2 \ell _3^2 \left(\kappa  x_1^2-2\right)}{\left(\ta^2+\kappa  x_1^2-3\right)^2}+r^2}}\, , \hspace{4.3cm}
\end{equation}
\begin{equation}
g_{t\phi} = -\frac{4 \ta \ell _3 \left(\ta^2-\kappa  x_1^2+1\right)}{\left(3-\ta^2-\kappa  x_1^2\right)^2}\left(1+\frac{2 \sqrt{1-\ta^2} \nu  \ell _3}{\left(3-\ta^2-\kappa  x_1^2\right) \sqrt{\frac{4 \ta^2 \ell _3^2 \left(\kappa  x_1^2-2\right)}{\left(\ta^2+\kappa  x_1^2-3\right)^2}+r^2}}\right)\, , \;
\end{equation}
\begin{equation}
\begin{split}
g^{rr} = H(r) = & \, \frac{r^2}{\ell _3^2} - \frac{8 \left(1-\ta^2\right)^{3/2} \nu  \ell _3 \left(\ta^2-\kappa  x_1^2+1\right) \sqrt{\frac{4 \ta^2 \ell _3^2 \left(\kappa  x_1^2-2\right)}{\left(\ta^2+\kappa  x_1^2-3\right)^2}+r^2}}{r^2 \left(3-\ta^2-\kappa  x_1^2\right)^3} \hspace{0.4cm}\\
&+\frac{16 \ta^2 \ell _3^2 \left(\ta^2-\kappa  x_1^2+1\right)^2}{r^2 \left(\ta^2+\kappa  x_1^2-3\right)^4}+\frac{4 \left[\left(\ta^2+1\right) \kappa  x_1^2-4 \ta^2\right]}{\left(\ta^2+\kappa  x_1^2-3\right)^2}\, .
\label{Hrparameters}
\end{split}
\end{equation}

\section{Particular case: classical BTZ black hole}
\label{App:ClassicalBTZ}

The classical BTZ black hole is described by the line element
\begin{equation}
    ds^2=-N(r)^2 \dd t^2 + N(r)^{-2}\dd r^2 + r^2 (N^\phi(r) \dd t + \dd\phi)^2\,,
\end{equation}
where the squared lapse function reads
\begin{subequations}
\begin{align}
    &N(r)^2 = -8{\cal G}_3 M + \frac{r^2}{\ell_3^2} + \frac{(4{\cal G}_3 J)^2}{r^2}\,,\\
    &N^\phi(r) =-\frac{4{\cal G}_3 J}{r^2}\,.
\end{align}
\end{subequations}
This metric is recovered in the $\nu\to0$ limit of the expressions provided in \cref{App:MetricFunctionsqBTZ}, after using also Eqs.~\eqref{M} and~\eqref{J}.\footnote{Note that the normalization of the mass and angular momentum in these expressions does not match  those of the usual BTZ metric~\cite{Banados:1992wn,Banados:1992gq}, but they are harmonious with the definitions adopted for the qBTZ metric in~\cite{Emparan:2020znc}. The difference can be attributed to the use of the renormalized Newton constant ${\cal G}_3$.}

The extremality condition for the classical BTZ spacetime is obtained by demanding that a degenerate root of the lapse function occurs at the horizon $r_\text{H}$. This yields the following system:
\begin{subequations}
\begin{align}
    &-8{\cal G}_3 M + \frac{\rH^2}{\ell_3^2} + \frac{(4{\cal G}_3 J)^2}{\rH^2} = 0\,,\\
    &\;\;\;\;\frac{\rH}{\ell_3^2} - \frac{(4{\cal G}_3 J)^2}{\rH^3} = 0\, ,
\end{align}
\end{subequations}
whose solution is $|J| = \ell_3 M$ and $\rH^2 = 4\ell_3{\cal G}_3 |J| = 4 \ell_3^2{\cal G}_3 M$.
It was shown in~\cite{Rocha:2011wp} that this extremal bound cannot be surpassed with test particles if the initial black hole is extremal.

\section{Computation of $\delta H$}
\label{App:deltaHzero}

Instead of considering Eq.~\eqref{Hrparameters}, which is more complicated, we take a step back and consider Eq.~\eqref{grrgeneral} instead. This time, however, we do not specify the form of the last term encoding backreaction effects, but we simply treat it as a function $F$ of $\ta$, $x_1$ and $r$. We thus consider $g^{rr}$ in the form
\begin{equation}
    H(r) = \frac{r^2}{\ell_3^2}-8 M + \frac{\left(4J \right)^2}{r^2} + F(r, \ta, x_1)\, ,
    \label{Happendix}
\end{equation}
where, of course, $M = M(\ta, x_1)$ and $J = J(\ta, x_1)$ (see Eqs.~\eqref{M} and~\eqref{J}).
In this appendix we shall absorb factors of ${\cal G}_3$ in $M$ and $J$ to reduce cluttering the equations.

We now evaluate $\delta H$ according to Eq.~\eqref{deltaH} and compute it at the minimum. Again, only the variations with respect to $\ta$ and $x_1$ survive. According to Eq.~\eqref{Lbound}, we simply express $L_\text{max} = \ell_\text{max}(\rH, \ta, x_1) \, E$. The explicit form of $\ell_\text{max}$ will not be needed. 

The variation of Eq.~\eqref{Happendix} reads
\begin{equation}
\begin{split}
    \delta H &= \left(-\ta^2+x_1^2+3\right)^2 \left[(1-\ta^2) \ell _3 \left(3-\ta^2-x_1^2\right) \left(\ta^2+x_1^2+1\right) \right]^{-1}\\
    &\biggl\{\frac{\delta M\left[\ell _3 \left(\ta^4+\left(6 \ta^2+4\right) x_1^2+12 \ta^2+x_1^4+3\right)-\ta \left(\left(\ta^2+9\right) x_1^2+4 \left(\ta^2+3\right)+x_1^4\right) \ell _{\max}\right]}{x_1} {\cal X} \\
    &-\Big[\delta M \left(\ta^4+\left(\ta^2+1\right) x_1^2+6 \ta^2-3\right) \ell _{\max}-2 \ta \, \delta M \ell _3 \left(3 \ta^2+x_1^2-1\right)\Big]{\cal Y} \biggr\}\,,
\end{split}
\label{deltaHexplicitappendix}
\end{equation}
where
\begin{equation}
{\cal X} = \partial_{x_1}F - 8\partial_{x_1} M +\frac{32 J\, \partial_{x_1}J}{r^2} 
\qquad \text{and} \qquad
{\cal Y} = \partial_{\ta} F - 8 \partial_{\ta} M +\frac{32 J \, \partial_{\ta} J}{r^2} \,.
\end{equation}

Every function of $r$ is understood to be computed at the minimum $r = r_\text{min}$. If this minimum coincides with an horizon (namely, in the extremal case), we can invert Eq.~\eqref{Happendix} to get $F(\rH, \ta, x_1)$:
\begin{equation}
    F(\rH, \ta, x_1 ) = 8M -\frac{\rH^2}{\ell_3^2}-\frac{\left(4J \right)^2}{\rH^2}\, .
\end{equation}

We see that $\partial_{x_1} F$ and $\partial_{\ta} F$ are such that the combinations ${\cal X}$ and ${\cal Y}$ appearing in~\eqref{deltaHexplicitappendix} vanish. Therefore, this confirms that $\delta H = 0$ when the background is an extremal state. This also explains the generality of the result: it does not depend on $\nu$ since its information is completely encoded in $F$. The result is also independent of $\kappa$, since it appears only in the explicit forms of $M$, $J$ and $F$.

Finally, the steps above show that, at linear order in the perturbations, the explicit form of the angular momentum of the incoming particle is irrelevant when considering the extremal configuration. In this case, Eq.~\eqref{Lbound} is a condition only for the particles to be absorbed, but it has no consequences on the net result. 

\end{appendix}

\bigskip
\bibliography{refs}

\end{document}